\documentclass[journal, onecolumn, 10pt]{IEEEtran} 

\usepackage{graphicx}
\usepackage{tabularx}
\usepackage{amsmath}
\usepackage{bm}
\usepackage{amssymb}
\usepackage[square,sort,comma,numbers]{natbib}
\usepackage{color,soul}
\usepackage{xcolor}
\usepackage{subcaption}
\usepackage{mathtools}
\usepackage{bbm}
\usepackage{ mathrsfs }
\usepackage{hyperref}
\hypersetup{backref, colorlinks=true, linkcolor=blue, citecolor=blue, urlcolor = blue}

\newcommand\numberthis{\addtocounter{equation}{1}\tag{\theequation}}

\newcommand{\subsc}[1]{\ensuremath{_{\textrm{#1}}}}

\newcommand{\PEPin}{$\pi_E(p\subsc{in})$}
\newcommand{\dmin}{$d\subsc{min}$}
\newcommand{\theoremTitle}{Proposition }
\newtheorem{theorem}{\theoremTitle}
\newcommand{\VersionLength}{long}
\providecommand{\verlong}{\ifthenelse{\equal{\VersionLength}{long}}}
\newcommand{\VersionCols}{single}
\providecommand{\dcol}{\ifthenelse{\equal{\VersionCols}{double}}}

\newif\ifshowtodo
\showtodofalse

\begin{document}

\title{Two-layer Coded Channel Access with Collision Resolution: Design and Analysis  
}

\author{
	
	\IEEEauthorblockN{
		MohammadReza Ebrahimi\IEEEauthorrefmark{1},
		Farshad Lahouti\IEEEauthorrefmark{2},
		Victoria Kostina\IEEEauthorrefmark{2}
	}

	\IEEEauthorblockA{
		\IEEEauthorrefmark{1}Department of Electrical and Computer Engineering,
		University of Toronto, Ontario, Canada
	}\\
	\IEEEauthorblockA{
		\IEEEauthorrefmark{2}Electrical Engineering Department, California Institute of Technology, USA
	}

	Emails: mr.ebrahimi@mail.utoronto.ca, lahouti@caltech.edu, vkostina@caltech.edu
    \thanks{This work was supported in part by the National
    Science Foundation (NSF) under grant CCF-1817241. A preliminary report on this research was presented at ISIT'17 \cite{ebrahimi2017coded}.
    
    This work has been submitted to the IEEE for possible publication.  Copyright may be transferred without notice, after which this version may no longer be accessible.
    }	 
}

\maketitle

\begin{abstract}
We propose a two-layer coding architecture for communication of multiple users over a shared slotted medium enabling joint collision resolution and decoding. Each user first encodes its information bits with an outer code for reliability, and then transmits these coded bits with possible repetitions over transmission time slots of the access channel. The transmission patterns are dictated by the inner collision-resolution code and collisions with other users' transmissions may occur. We analyze two types of codes for the outer layer: long-blocklength LDPC codes, and short-blocklength algebraic codes. With LDPC codes, a density evolution analysis enables joint optimization of both outer and inner code parameters for maximum throughput. With algebraic codes, we invoke a similar analysis by approximating their average erasure correcting capability while assuming a large number of active transmitters. The proposed low-complexity schemes operate at a significantly smaller gap to capacity than the state of the art. Our schemes apply both to a multiple access scenario where number of users within a frame is known a priori, and to a random access scenario where that number is known only to the decoder. In the latter case, we optimize an outage probability due to the variability in user activity. 

\end{abstract}

\IEEEpeerreviewmaketitle

\section{Introduction}

Random access (RA) protocols are at the heart of many modern cellular standards for contention resolution over their control channels or for medium access control in wired or wireless local area networks. Almost all these protocols are developed based on variations of the legacy (slotted) Aloha (SA) or carrier sense multiple access. These legacy protocols,  while having served us well, do not meet the requirements in the emergent machine-to-machine communications and the Internet of things with limited communication delays or a massive number of communicators. Indeed, random access protocols are to be reinvented. 

Efforts on improving the performance of SA methods were revived by contention  resolution  diversity  slotted  Aloha (CRDSA) \cite{CRDSA}, which replaces the concept of destructive collisions in SA by successive interference cancellation (SIC) strategies. Specifically, each user is prescribed to send its packet twice over two time-slots of a single contention frame. The receiver then detects the signals received without a collision, and subsequently removes the associated interference from time-slots in which their replicas were transmitted. The process iteratively continues until all transmitted packets are recovered. In \cite{IRDSA, MDIRSA}, the SIC is represented by an irregular bipartite graph and the users adopt repetition rates from a given distribution for the transmission of each packet. Utilizing a density evolution analysis, this distribution is optimized to maximize throughput. Other improvements are presented in e.g. \cite{coded_SA,paolini2011graph,paolini2015coded,rateless,IRSA_practical}. Many of the works in this domain assume an error free transmission, and the loss is only due to collisions that are not resolved. Since errors in the recovery of one packet will propagate and harm the SIC process, the error free assumption is typically ensured using long and low-rate channel codes. 

There are works with similar spirit in the context of multiple access channels, where message passing algorithms are utilized to enable multiuser detection. This is indeed the case in sparse coded multiple access \cite{SCMA, nikopour2014scma}, where each user is provided with a sparse codebook and the codewords of different users are transmitted over shared orthogonal resources, e.g., OFDM tones. In a similar scenario, \cite{interplay} rigorously examines the role of error correcting codes in iterative decoding and multiuser detection. A multiple access code based on analog fountain codes \cite{AFC_near_cap} purposely designed for wireless fading channels is proposed in \cite{AFC}. Contemporaneously with the conference version of the current work \cite{ebrahimi2017coded}, a two-layer coding scheme for collision resolution and noise removal was proposed in \cite{ordentlich2017mac},  in which a collision resolution code to resolve $N$ collisions is constructed from the columns of an $N$-error correcting BCH code. A specific scenario following the two-layer architecture known as the two-user joint iterative decoding, is discussed in \cite{Amraoui2002, Roumy2007, palanki2001graph} for Gaussian and binary adder multiple access channels. 

In this paper, we present a two-layer coded channel access architecture for communications over binary adder access channel with erasures (Section \ref{sec_proposed_ra}). The inner layer is meant to resolve collisions, while the outer layer targets erasures; the layers operate in lockstep while passing messages back and forth, resulting in a joint resolution of collisions and erasures. The inner layer is described by a transmission pattern to transmit the coded bits with repetitions according to an optimized probability distribution over a frame. 
In principle, one may use any point-to-point error correcting code as the outer erasure-correcting code in our two-layer architecture. In this paper, we study two possibilities for the outer layer code: a low-density parity check (LDPC) code drawn from an ensemble of LDPC codes (Section \ref{sec_assymptotic_LDPC}), and a short algebraic code from a family of such codes (Section \ref{sec_assymptotic_classic}).  Density evolution analyses presented in Sections \ref{sec_assymptotic_LDPC} and  \ref{sec_assymptotic_classic} enable an optimization of the code parameters in both layers subject to an imposed complexity constraint. The analysis of the short codes in Section \ref{sec_assymptotic_classic} is aided by an efficient approximation of the decoding performance of the algebraic code using its parity check matrix. 
Both a two-layer coded multiple access (TCMA) scenario where the number of users within a frame is known at the transmitters and the receiver, and a two-layer coded random access (TCRA) scenario, where this information is only available at the receiver, are considered. In the latter case, the code is designed to achieve a certain probability of outage induced by the randomness in the number of users. Simulation results in Section \ref{sec_results} demonstrate that TCRA with large-blocklength LDPC codes at the outer layer (TCRA-LDPC) halves the gap to the outage capacity compared to the performance of an equivalent perfect orthogonal transmission scheme (for an average of 6 active users) and noticeably outperforms a benchmark due to prior art, in which the two layers are designed and operate separately. As expected, the TCRA with algebraic codes of short blocklength at the outer layer underperforms TCRA-LDPC, however, it still outperforms the prior art benchmark while maintaining a low application level delay and payload size requirement.  

Compared to the scheme in \cite{Amraoui2002, Roumy2007, palanki2001graph}, TCMA introduces an additional degree of freedom in system design by enabling a joint optimization of the code and transmission patterns with repetitions, achieving wider design trade-offs for complexity, user/system rate and performance. In  \cite{Amraoui2002, Roumy2007, palanki2001graph}, the transmission pattern is fixed and each coded bit is transmitted only once over bit intervals. Compared to the traditional coded CDMA schemes, which operate with a given bandwidth expansion and a spreading sequence for each coded bit, with all users colliding with each other at all times,  both layers of our TCMA scheme are jointly optimized, and not all users may necessarily collide over a time slot. 
Compared to the conference version of this paper \cite{ebrahimi2017coded}, the results on short algebraic codes, as well as those on the multiple access scenario with an a priori known number of users, are new. 

\section{System Model and Assumptions} \label{sec_system_model}

\subsection{Channel Access Model}
We consider a network consisting of one receiver (base station) and a total of $N$ transmitters. The time is divided into (contention) frames of length $T_f$ that are composed of $T$ bit intervals of length $T_b$, hence $T_f=T*T_b$. For the sake of simplicity, we assume that each active transmitter sends a packet of size $k$ information bits within the $T$ bit intervals ($k\leq T$) of each frame, leading to a common transmission rate of $R_{t} = k/T$ bits per user per bit interval. Each transmitter first encodes its $k$-bit packet into $n$ bits ($k  \leq n$) and then transmits them over $T$ bit intervals $(n \leq T)$ with possible repetitions according to a random but fixed pattern. Different transmitters can use the same bit interval, colliding with each other, however, self-collisions are not allowed, that is, the same transmitter can at most send one bit over a bit interval. 

Frame synchronization for channel access management is typically facilitated by the transmission of beacon messages by the receiver to mark the start of a frame.
In a multiple access channel (MAC) scenario, the number of active users within a frame is assumed known at the transmitters. In a random access channel (RAC) scenario, the transmitters are only aware of the start and the end of a new contention frame, as well as the codes to use. Therefore, they are unaware of the real number of active users. We assume that the receiver knows both the set of active transmitters and their transmission patterns (location of their bits) in the frame. These transmission patterns may be agreed upon between the transmitters and the receiver in a setup phase. Ideas on practical ways to handle such issues in a RAC are presented at the end of Section \ref{sec_results}, after more is discussed about the decoding procedure.

To model variable user activity in a RAC, we assume that each transmitter participates in a frame with the probability $p_a$. Under the assumption of a large number of transmitters $N$ and a low activity probability $p_a$, the number of active transmitters in a frame, $N_a$, closely follows a Poisson distribution with parameter $Np_a$. We consider bit intervals and bit transmissions in each interval in this paper, however, the scheme can be generalized to time-slots and bit sequences.

\subsection{Channel Model}

We consider a binary adder random access channel with erasures, whose inputs $x_i \in \{0, 1\}$, $1 \leq i \leq N_a$ are binary, and whose output $r\in\{0,1,2,\ldots,N_a\}\cup\{e\}$ is given by
\begin{equation} \label{eq_BARAC}
r = 
\begin{cases}
\sum\limits_{i=1}^{N_a} x_i,& \text{with probability } 1 - \epsilon\\
e,              & \text{with probability } \epsilon
\end{cases}, 
\end{equation} 
where, as before, $N_a$ is the number of active transmitters in a frame and $e$ designates an erasure.
This simple channel models the situation in which a digital encoder and decoder communicate over an analog channel via modulator/demodulator; the modulator converts bits to analog signals under a power constraint; the demodulator, having observed a sum of the signals from all the users and a random thermal noise, digitizes it by simple quantization. If the demodulator deems the received signal too noisy to digitize reliably, it declares an erasure and passes it on to the coding layer to resolve. Specifically, for BPSK in high SNR regime with $N_a$ users, the output of the demodulator $\bar{r}$ is in $\{-N_a, -N_a+2, \hdots, N_a-2, N_a\}$. A simple transformation $r=(\bar{r}+N_a)/2$ converts the received signal in each slot to $r\in\{0,1,2,\ldots,N_a\}$. In the presence of a noise, the demodulator declares an erasure when it cannot reliably detect the received symbol; hence we have $r\in\{0,1,2,\ldots,N_a\}\cup\{e\}$. 

Regardless of the number of active users $N_a$, due to the symmetry of the channel in \eqref{eq_BARAC}, its capacity-achieving distribution is Bernoulli(1/2), and the sum-rate capacity (for $N_a$ active users) is
\begin{align}\label{EqCNa}
    C_{N_a} &=  (1 - \epsilon)\left(N_a - \frac 1 {2^{N_a}} \sum_{i = 0}^{N_a} {N_a \choose i} \log_2 {N_a \choose i} \right).
\end{align}
This is readily obtained by computing the mutual information between the vector of inputs and the output for the capacity achieving distribution, similar to what is reported in \cite{chang1979codingforTuser} for the noiseless adder multiple access channel.

If the number of active users $N_a$ is random but stays constant within a contention frame, the \emph{outage capacity} is an appropriate performance measure: 
\begin{equation}
    C_{p} = \max \left\{ R_t \colon \Pr \left[C_{N_a} < N_a R_t\right] \leq p \right\}, \label{eq_Cout}
\end{equation}
where 
$p$ is the \emph{outage probability}, i.e., the probability that the system cannot support the target rate $R_t$ given the randomness in the number of active users $N_a$. This metric is used for design and evaluation of access schemes for RAC in the sequel.
 
\section{Two-layer Coded Channel Access Scheme} \label{sec_proposed_ra}

In this section, we describe our two-layer coding architecture for medium access, an inner layer dealing with collisions and an outer layer dealing with erasures. During the decoding procedure, these two layers run iteratively and update each other, enabling joint contention resolution and decoding of messages.  

In the rest of this section, we first introduce the encoder/transmitter model and the system parameters, then we discuss the decoder/receiver.

\subsection{Encoder/Transmitter} \label{subsec_enctx}

\begin{figure}[hb!]
	\begin{center}
		\includegraphics[height=10.5cm, keepaspectratio]{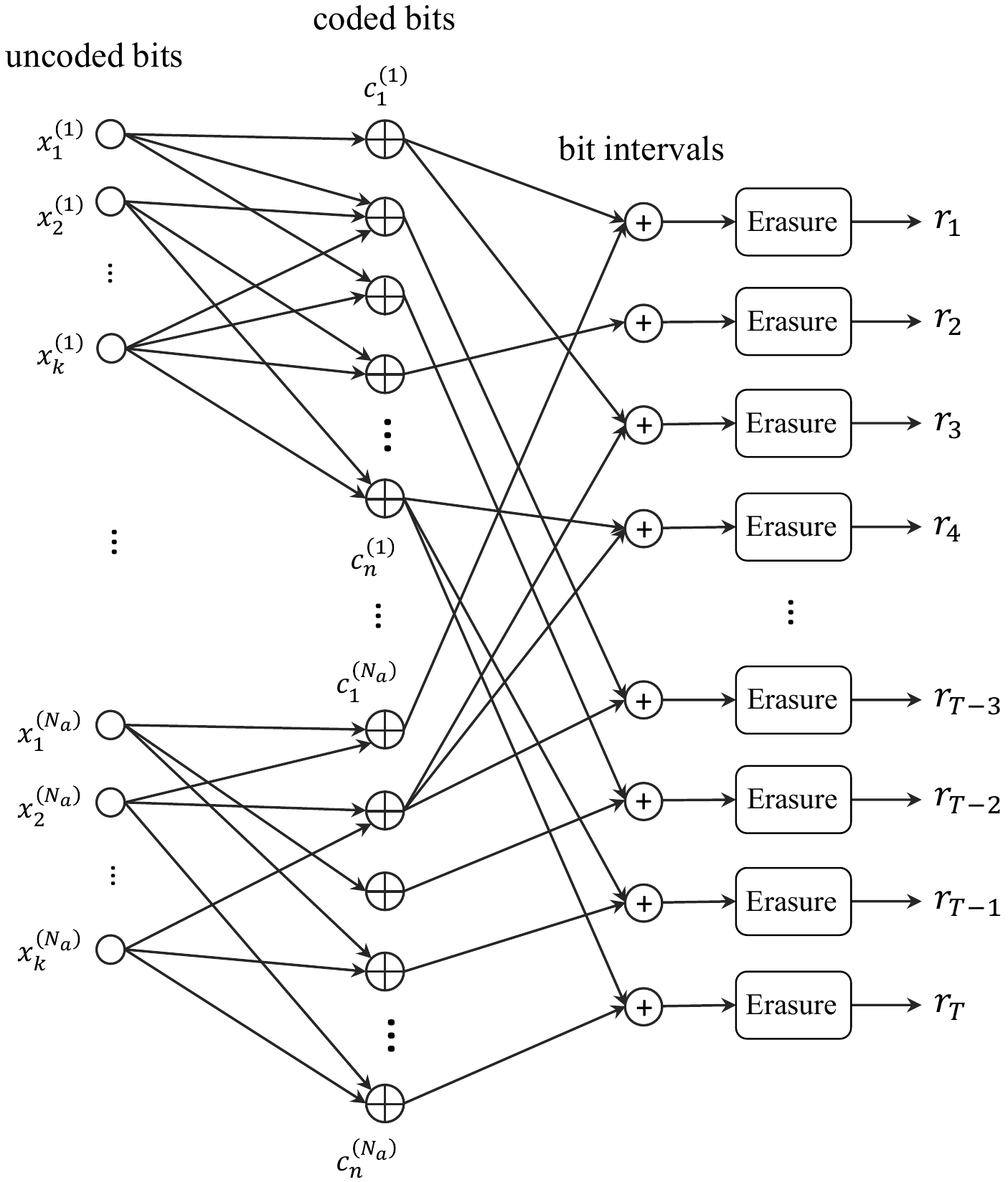}
	\end{center}
	\caption{Transmission diagram of the two-layer coded medium access system: $k$ information bits of each of the $N_a$ transmitters are first mapped into $n$ coded bits. These coded bits are then transmitted with possible repetition across the bit intervals of the contention frame. The adder-erasure channel \eqref{eq_BARAC} adds the bits that share the same bit interval, possibly introducing erasures and producing the outcomes $r_1, \ldots, r_T$ for the receiver to decode.}
	\label{fig_encoder}
\end{figure}

Fig. \ref{fig_encoder} shows the encoding diagram of the proposed coded medium access scheme. First, using an error correcting code of rate $R_{\mathsf o }=k/n$, the $j$-th transmitter encodes its $k$-bit packet $\bm{x}^{(j)}$ to create $n$ coded bits $\bm{c}^{(j)}$. To each coded bit $c_i^{(j)}\in\{0,1\}, 1\leq i \leq n$ there corresponds a repetition rate $d_i^{(j)}$ and a repetition pattern indicating which $d_i^{(j)}$ out of the $T$ bit intervals will carry this bit. Each transmitter uses a bit interval at most once in a frame. 

The common per-user rate $R_t = k /T$ can be expressed in terms of outer erasure-correcting code rate $R_{\mathsf o } = k / n$ and inner collision-resolution code rate $R_{\mathsf i} = n / T$ as
\begin{align}
	R_t = R_{\mathsf o } R_{\mathsf i}.
\end{align}
Of course, the rate of any concatenated code is equal to the product of the rates of its component codes. From \eqref{eq_BARAC}, it is clear that the received value on an interval is the summation of coded bits sent on that interval from different transmitters, which could be erased with probability $\epsilon$.

In our analysis, we assume that the repetition rates $d_i^{(j)}$ are generated according to the probability mass function $\{\Gamma\}$, and that the repetition patterns are randomly selected without biasing any particular location. These choices are at the purview of code design and are fixed during transmission. For simplicity, we consider the case where all the transmitters use an identical erasure-correcting outer layer code, but the repetition patterns are generated individually for each user.  

\subsection{Decoder/Receiver} \label{subsec_decrx}

\begin{figure}[tb!]
    \centering
	\begin{subfigure}[b]{0.24\textwidth}
		\includegraphics[width=\linewidth]{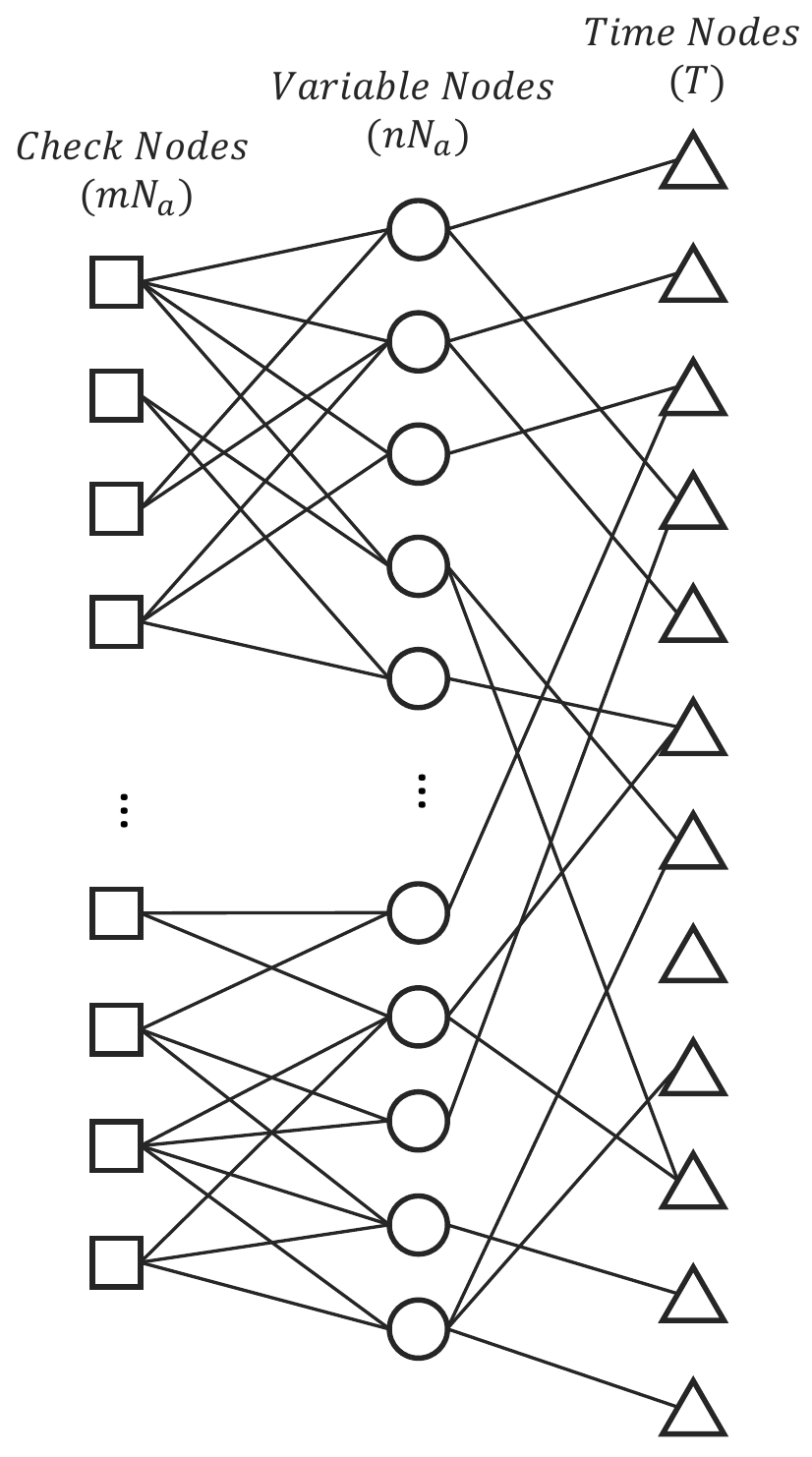}
		\caption{} \label{fig_decoder_a}
	\end{subfigure}
	\hspace*{5em} 
	\begin{subfigure}[b]{0.21\textwidth}
		\includegraphics[width=\linewidth]{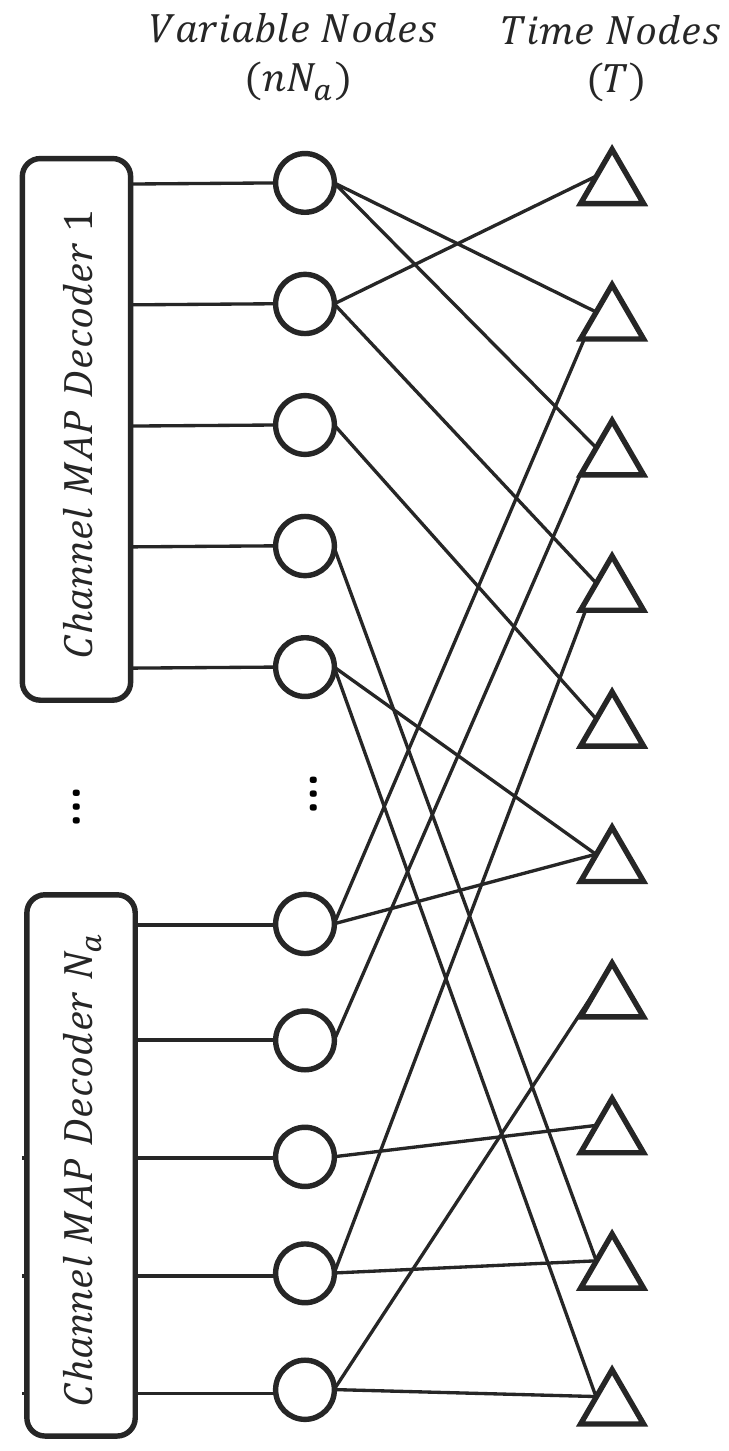}
		\caption{} \label{fig_decoder_b}
	\end{subfigure}
	\caption{Decoding graph with the outer layer implemented using \subref{fig_decoder_a} LDPC codes with message passing decoder, \subref{fig_decoder_b} linear block codes with block MAP decoder.} \label{fig_decoder}
\end{figure}

Message passing is performed on the decoder's factor graph to recover the value of the bit (variable) nodes. Fig. \ref{fig_decoder} shows the factor graph of the receiver with \subref{fig_decoder_a} LDPC codes with message passing decoding, and \subref{fig_decoder_b} short algebraic codes with MAP decoding at the outer layer.  Messages passed on the edges of the decoder's graph are either $e$, indicating that the value of the variable node has not been recovered yet, or 0 / 1, reflecting the value of the associated variable node. 

For each node, we define a node-perspective and an edge-perspective degree distribution. The former determines the fractions $\{A_0,A_1,\cdots,A_d\}$ of nodes with degrees $0$ to $d$, while the latter specifies the fractions $\{a_1,a_2,\cdots,a_d\}$ of edges connected to nodes with specified degrees. We use the standard polynomial representations $A(x)=\sum_{i=0}^{d}A_i x^i$, $a(x)=\sum_{i=1}^{d} a_i x^{i-1}$ for these degree distributions. Note that the edge-perspective polynomial is obtained by the normalized derivative of the node-perspective polynomial. Table \ref{table_degreedists} summarizes the notations for the edge and node-perspective degree distributions for the different layers of the decoding graph. 

\begin{table}[t!]
    \centering
	\caption{Degree Distributions of the Decoding Graph 
}
	\label{table_degreedists}
	\begin{tabular}{|c|m{24em}|}	
		\hline
		\bfseries Symbol & \bfseries Degree Distribution (Probability Mass Functions) \\
		\hline
		\hline
		$ \{\lambda\} / \{\Lambda\} $ & Variable edge/node-perspective degree dist. in outer layer \\
		\hline
		$ \{\rho\} / \{P\} $ & Check edge/node-perspective degree dist. \\
		\hline
		$ \{\gamma\} / \{\Gamma\} $ & Variable edge/node-perspective degree dist. in inner layer \\
		\hline
		$ \{\psi\} / \{\Psi\} $ & Time edge/node-perspective degree dist. \\
		\hline
	\end{tabular}
\end{table}

At each node, the processing rule generates an output message to a neighbor node, based on the messages it received from its other neighboring nodes.
At a time node of degree $d$ with input messages $m_i$, $1\leq i \leq d-1$, and a received value of $r$ from the channel, the processing rule to generate the message to a variable node is:

\begin{equation} \label{eq_tnode}
f_T (r, m_1, m_2, ..., m_{d-1}) = 
\begin{cases}
0, & r = \sum\limits_{i:m_i\neq e} m_i\\
1, & r = 1 + \sum\limits_{i:m_i\neq e} m_i + \sum\limits_{i:m_i=e} 1 \\
e, & \text{otherwise.}
\end{cases}
\end{equation}
Equation \eqref{eq_tnode} states that a time node can recover the unknown bits whenever there is no other choice for them except being all zeros or all ones. Note that the message sent from an erased time node is always $e$.

For the outer layer, depending on the application, LDPC codes or short linear block codes can be used.

\subsubsection{Using LDPC Codes}

Fig. \ref{fig_decoder}\subref{fig_decoder_a} shows the factor graph \cite{kschischang2001factor} of the decoder, which is obtained by combining $N_a$ Tanner graphs of the senders' LDPC codes to the transmission graph of the coded bits. Both layers use message passing decoding on bipartite graphs. Check nodes of the LDPC Tanner graphs (designated by squares) and time nodes (triangles) are the function nodes, and bit nodes (circles) are the variable nodes in this graph.

The $N_a$ LDPC decoders behave exactly as the standard erasure decoders, i.e., the output message of a check node equals the modulo-2 sum of its input messages if none of them is $e$, and $e$ otherwise: 

\begin{equation} \label{eq_cnode}
    f_C (m_1,m_2, ..., m_{d-1}) = 
    \begin{cases}
        \bigoplus\limits_{i=1}^{d-1} m_i, &   \forall i \in \{1,2,...,d-1\}: m_i \neq e \\
        e, & \text{otherwise}
    \end{cases}
\end{equation}

where $\bigoplus$ denotes modulo-2 sum. Since the erasure channel does not change received values but erases them, the non-erased input messages to a variable node always agree. Therefore, when the value of a variable node is recovered from a time node or a check node equation, it always equals the real value sent by the transmitter. As a result, the outgoing message from a variable node is $e$ if all the input messages are erased; otherwise, it equals the non-erased value of the input messages:

\begin{equation} \label{eq_vnode}
	f_V(m_1, m_2, ..., m_{d-1}) = 
	\begin{cases}
		m_i,	& \exists i \in \{1,2,...,d-1\}\colon m_i \neq e \\
		e,	& \text{otherwise}
	\end{cases}
\end{equation}

Note than even with just one non-erased input from either the time nodes or the check nodes in any step of the iterative decoding, the true value of the variable node is resolved and accordingly could be used for all the outgoing messages from the variable node.

\subsubsection{Using Short Algebraic Codes}

Fig. \ref{fig_decoder}\subref{fig_decoder_b} shows the factor graph of the receiver that uses $N_a$ individual decoders for the erasure-correcting outer layer. The  decoding architecture  in  general  can accommodate  various  types  of  the  individual  decoders  for  the  erasure-correcting  linear  codes. For the analysis and design, however, we focus on block MAP decoders in the sequel, whereby depending on the erasure pattern of the input codeword, the decoder either recovers all the erased bits or fails to recover any of them.

\subsubsection{Design, Operation and Complexity Considerations}
As the transmitters select bit intervals according to their repetition patterns, the time node degree distribution $\{\Psi\}$ cannot be designed directly and is obtained from repetition rate distribution of variable nodes, i.e., $\{\Gamma\}$. As we shall see, this is essential in the analysis and design of the decoders performance in the sequel.

The probability that the $j$th transmitter selects a bit interval is $d^{(j)}/T$, where $d^{(j)} = \sum_{i=1}^{n} d_i^{(j)}$ is the total number of transmissions for sender $j$. As the code length grows, according to the law of large numbers,
\begin{equation}
	\frac{1}{n}d^{(j)} \rightarrow \mathbb E \left\lbrace  d_i^{(j)} \right\rbrace = \sum_i{i \Gamma_i} = \Gamma'(1),
\end{equation}
where $\Gamma(.)$ is the polynomial representation of the repetition rate distribution. Hence, the degree of time nodes follows the binomial distribution
\begin{equation}
	\{\Psi\} \sim  B \left( N_a, \frac{n}{T}\Gamma'(1) \right) = B \left( N_a,R_{\mathsf i}\,\Gamma'(1) \right).
\end{equation}

With LDPC codes, the decoding complexity is determined by the total number of edges in the decoder's graph, as the number of operations at each node has a linear relationship with its degree. The number of edges in the graphs of the outer and the inner layers are $nN_a\sum_i i\Lambda_i$ and $nN_a\sum_i i\Gamma_i$, respectively. Hence, the complexity of decoding in one iteration is $O\left( nN_a\sum_i i(L_i+\Gamma_i)\right) = O\left( nN_a (\Gamma'(1) + \Lambda'(1) \right) $. The complexity remains linear in terms of the number of active users, code length and maximum degree of nodes. If short linear codes are used at the outer layer, the decoding complexity depends on that of the individual decoders. Still the share of the complexity due to the inner layer remains the same. Therefore, the decoding complexity grows linearly with the number of users in both cases.

Introducing a schedule will finalize the definition of the decoder. The decoder starts a decoding iteration with the collision resolution inner layer: time nodes send messages containing their estimates \eqref{eq_tnode} of the values of the variable nodes to the set of variable nodes. The variable nodes update and broadcast their values \eqref{eq_vnode} to the set of check nodes (case of LDPC) or the MAP decoders (case of short codes). The check nodes or the MAP decoders then report back to the variable nodes. Finally, the variable nodes send their updated values \eqref{eq_tnode} to the time nodes, concluding a decoding iteration. The decoder continues until either no new variable node is recovered during an iteration, or a maximum number of iterations is reached. Note that this schedule is motivated by the particular processing rule at variable nodes described in \eqref{eq_vnode}.


\section{Asymptotic Analysis of Decoding Threshold - LDPC Codes} \label{sec_assymptotic_LDPC}

Like other message-passing decoders, our decoder exhibits threshold behaviors. The threshold is determined both by the channel erasure probability and users transmission rate. More accurately, under the assumption of asymptotically large code and frame length, if the number of active transmitters does not exceed $N_a^*$ and the erasure probability is less than $\epsilon^*$, the receiver is able to recover message bits completely as long as the per-user rate is less than the threshold, $R_t^*(N_a^*,\epsilon^*)$. An equivalent characterization of the decoding threshold is the maximum number of users sustainable by the system that is compatible with a given rate and erasure probability, $N_a^*(R_t^*,\epsilon^*)$.
We use density evolution \cite{LDPC_richardson} to calculate the exact average performance of a code ensemble in each iteration under the assumption of infinite code length. 
In asymptotic regime, the decoding graph becomes a tree, which implies the independence of incoming messages to a node. Therefore, as customary, density evolution is analyzed on an idealized natural schedule defined on a tree \cite[p.82]{modern_coding_theory} instead of analyzing the efficient decoding schedule introduced before for the non-asymptotic setting. Specifically, with the natural schedule, the message from a variable node to any single check (time) node is a function of the messages it receives from all neighboring time (check) nodes and the rest of the neighboring check (time) nodes. We compare the analysis results for obtaining the decoding threshold and the decoding simulation results in Section \ref{sec_results}. In the following, we first derive the erasure probability transfer function of each node in Section \ref{subsec_erasuretransfer_LDPC}. Then, we introduce a set of equations to recursively track the erasure probability in each iteration in Section \ref{subsec_trackdecoder}. Finally, we present optimized degree distributions in Section \ref{subsec_optimization}.

\subsection{Erasure Transfer Function of Nodes} \label{subsec_erasuretransfer_LDPC}
In this subsection, we find the probability $P\subsc{out}$ that the output message of a node equals $e$, given the erasure probability of input messages $p\subsc{in}$. Note that under the assumption of infinite code and frame lengths, all messages entering a node are independent, since there is no loop in the decoding graph of the receiver.

From the standard density evolution of erasure correcting LDPC codes, we obtain for the check and variable nodes:
\begin{align} 
P\subsc{out}{}_{,c} &= 1-\rho(1-p\subsc{in} ) \label{eq_cnode_tf}, \\
P\subsc{out}{}_{,v} &= \lambda(p\subsc{in}) \label{eq_vnode_tf}.
\end{align}

%
%
%

For the time nodes with the processing rule in \eqref{eq_tnode}, \theoremTitle \ref{theo_time_node_tf} presents the transfer function.
\begin{theorem} \label{theo_time_node_tf}
	Erasure transfer function of a time node is described by
	\begin{equation} 
		P\subsc{out}{}_{,t} 
		= 1-(1-\epsilon)\psi \left( 1-\frac{p\subsc{in}}{2} \right). \label{eq_tnode_tf}
	\end{equation}
\end{theorem}

\begin{IEEEproof}
	An examination of \eqref{eq_tnode} reveals that the unknown inputs can be recovered if all of them have the same value and $r$ is not erased, therefore the probability that a degree-$d$ time node with $q$ input erasures outputs an erasure is given by
	\begin{align*}
		P\subsc{out}{}_{,t|q,d} 
		&= 1 - (1-\epsilon) 2^{-q} 
	\end{align*}
		Assuming that $q$ follows a binomial distribution $B(d-1,p\subsc{in})$, we can compute the expectation over $q$ using its moment-generating function:
	\begin{align*} 
		P\subsc{out}{}_{,t|d} &= 1 - (1-\epsilon) \left( 1-\frac{p\subsc{in}}{2} \right) ^{d-1} .
	\end{align*}
Finally, averaging over $d$ yields 
	\begin{align*} 
		P\subsc{out}{}_{,t} &= 1 - (1-\epsilon) \sum_{i} \psi_i \left( 1-\frac{p\subsc{in}}{2} \right) ^{i-1} \\ 
		&= 1-(1-\epsilon)\psi \left( 1-\frac{p\subsc{in}}{2} \right).
	\end{align*}
\end{IEEEproof}

\subsection{Tracking Decoder Erasure Probability} \label{subsec_trackdecoder}

Having derived the erasure probability transfer function of the nodes, now we can track the erasure probability of a specific message along different iterations for fixed $N_a$ and $\epsilon$. This is done by stating the erasure probability of a message in terms of its value in the previous iterations. In the following, we denote by $ m_{ab}^{(\ell)} $ the message sent from the node $a$ to the node $b$ in the $\ell$th iteration. $P_{ab}^{(\ell)}$ is the probability that $ m_{ab}^{(\ell)} $ equals $e$, where $a$ and $b$ can stand for $v$ (for variable nodes), $c$ (for check nodes), and $t$ (for time nodes). Fig. \ref{fig_de_a} depicts the messages passed in one iteration. By expressing the erasure probability of each message in terms of the erasure probability of its generating messages, we reach a set of recursive equations for $P_{vc}^{(\ell)}$:

\begin{figure} [b!]
    \centering
	\begin{subfigure}[b]{0.4\textwidth}
		\includegraphics[width=\linewidth]{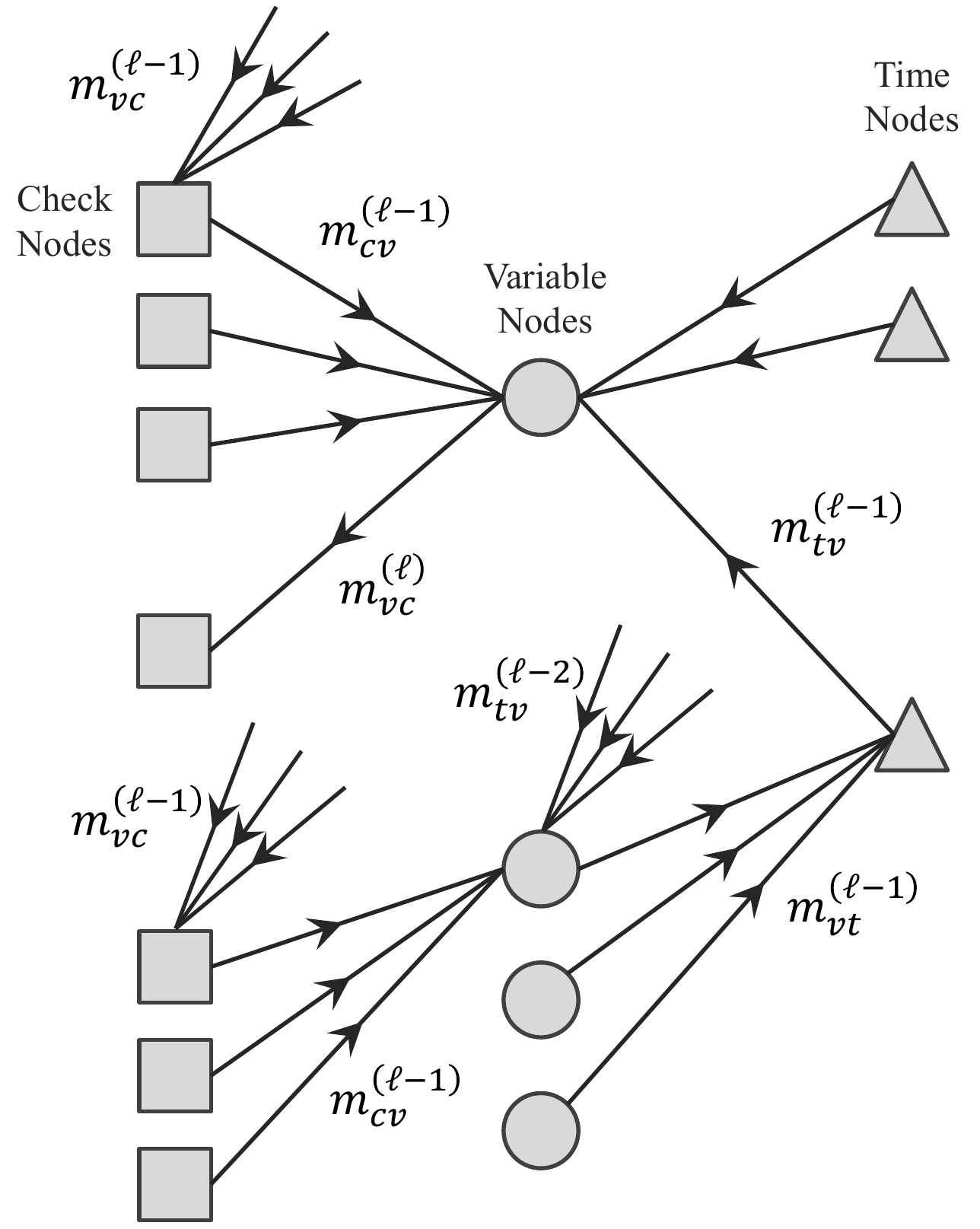}
		\caption{} \label{fig_de_a}
	\end{subfigure}
	\hspace*{5em} 
	\begin{subfigure}[b]{0.4\textwidth}
		\includegraphics[width=\linewidth]{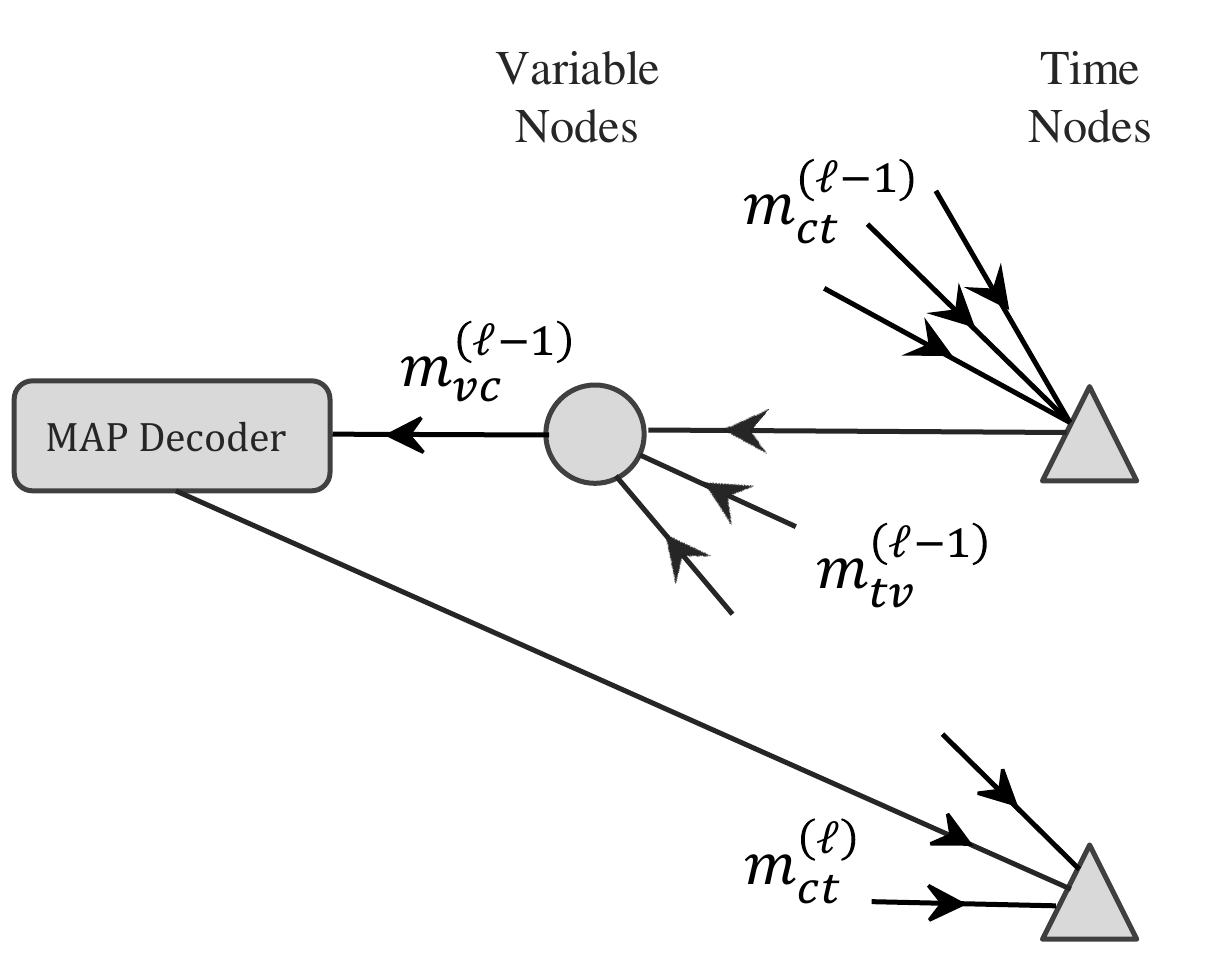}
		\caption{} \label{fig_de_b}
	\end{subfigure}
	\caption{Messages passed in one iteration with the outer layer using \subref{fig_de_a} LDPC codes with message passing decoder, \subref{fig_de_b} linear block codes with block MAP decoder.} \label{fig_de}
\end{figure}

\begin{align}
    & P_{vc}^{(\ell)} = \Gamma \left( P_{tv}^{(\ell-1)} \right) \lambda \left( P_{cv}^{(\ell-1)} \right) 
    \label{rec_1} \\
    & P_{tv}^{(\ell)} = 1 - (1-\epsilon) \psi \left( 1- {P_{vt}^{(\ell)}}/{2} \right)  
    \label{rec_2} \\
    & P_{vt}^{(\ell)} = \Lambda \left( P_{cv}^{(\ell)} \right) \gamma \left( P_{tv}^{(\ell-1)} \right) 
    \label{rec_3} \\
    & P_{cv}^{(\ell)} = 1 - \rho \left(  1-P_{vc}^{(\ell)} \right)
    \label{rec_4} \\
    & P_{vc}^{(0)} = 1 \label{rec_5} \\
    & P_{tv}^{(-1)} = 1 \label{rec_6}
\end{align}

Equations \eqref{rec_1} and \eqref{rec_3} are derived from \eqref{eq_vnode_tf}, while \eqref{rec_2} and \eqref{rec_4} are based on \eqref{eq_tnode_tf} and \eqref{eq_cnode_tf}, respectively. The definition of the initial values for the recursion in \eqref{rec_5} and \eqref{rec_6} are obvious since at the beginning of the decoding, every variable node is unknown and all edges carry erasure messages. Note that in each iteration, the fraction of non-recovered variable nodes is given by
\begin{equation} \label{eq_error_pr}
P_e^{(\ell)} =  \Gamma \left( P_{tv}^{(\ell-1)} \right) \Lambda \left( P_{cv}^{(\ell-1)} \right). 
\end{equation}

\subsection{Optimizing Degree Distributions} \label{subsec_optimization}
For maximum number of active senders $N_a^*$ and fixed channel erasure probability $\epsilon$, one can determine the threshold $R_t^*(N_a^*,\epsilon)$ of a given code ensemble $\{ \Gamma\}, \{ \lambda\}, \{ \rho\}$  by finding the largest value of $R_t$ which leads to $ \lim_{\ell \to \infty} P_{vt}^{(\ell)} = 0$ in recursion \eqref{rec_1}--\eqref{rec_5}. This could be done using a binary search on $R_t$. To look for the optimized degree distributions leading to the largest per-user rate threshold, we set up the following optimization problem. 

\begin{alignat*}{4}
	& \operatorname*{maximize}_{\{ \Gamma\}, \{ \lambda\}, \{ \rho\}} \quad && R_t^*(N_a^*,\epsilon) \numberthis \label{eq_opt} \\
	& \text{subject to:} 
	&& 0 \le R_{\mathsf o }(\lambda,\rho) \le 1 \\
	&	&& R_t^* \leq R_{\mathsf o } / \Gamma'(1) \\
	&	&& \Gamma_i = \lambda_i = \rho_i = 0, &&& i > d\subsc{max} \\
	&	&& \Gamma_i , \lambda_i , \rho_i \in [ 0,1 ], &&& i \leq d\subsc{max} \\
	&	&& \textstyle \sum_i \Gamma_i = \sum_i \rho_i = \sum_i \lambda_i = 1
	\label{eq_opt}
\end{alignat*}
The second constraint in \eqref{eq_opt} ensures the total number of transmissions per user does not exceed $T$, as users cannot use a bit interval more than once in a frame. We use differential evolution \cite{differential_evolution} to solve the above optimization problem. This meta-heuristic method, which is a combination of hill climbing and a genetic algorithm, is quite popular in code design problems, e.g. \cite{LDPC_richardson}, since it is capable of searching large parameter spaces effectively. It is particularly appealing in our case because it does not need the objective function to be explicitly stated.

For multiple access with a known maximum number of users $N_a^*$ and erasure probability $\epsilon$, the optimization \eqref{eq_opt} results in a maximum per-user rate. For random access, the design procedure to attain an acceptable outage probability $P\subsc{outage} = \Pr \left[N_a>N_a^*\right]$ is as follows. Knowing the distribution of the number of active users $N_a$, first the maximum number of senders $N_a^*$ compatible with $P\subsc{outage}$ is determined. Then, by solving \eqref{eq_opt} with $N_a^*$,  one can calculate the optimized degree distributions leading to the maximum per-user transmission rate.

Table \ref{table_G_result} presents the per-user rate thresholds corresponding to optimized degree distributions for target number of users $N_a^* = 7$ and different values of $\epsilon$ for $d\subsc{max}$ limited to $13$. As expected, as the probability of channel erasure grows, the code rate decreases to cope with channel impediments. 
Another observation from Table \ref{table_G_result} is the abundance of degree-2 variable nodes in the outer layer, as indicated by $\lambda(x) = x$, i.e. $\lambda_2=1$. Such a result is expected since rate threshold is the sole subject of maximization in \eqref{eq_opt}. The number of degree-2 variable nodes plays a major role in the trade-off between error floor and the threshold. For that matter, one can consider constraining $\lambda_2$ in \eqref{eq_opt} to trade rate threshold for error floor. An example of such a strategy is suggested in \cite{error_floor}. In the noiseless case ($\epsilon = 0$), the optimized code rate of $R_{\mathsf o } = 0.857$, which is the maximum possible under the complexity constraint $d_{\max} \leq 13$, falls short of $1$,  suggesting that in the noiseless scenario a better performance could be achieved by removing the outer layer. However, when the channel erasure exists, the collision-resolution layer alone cannot recover all the transmitted bits.
\theoremTitle \ref{theo_MacLowerBound} introduces a lower bound on the fraction of non-recovered bits for the collision-resolution layer.

\begin{table*}[t!]
	\caption{Optimized Degree Distribution for the Asymptotic Setting.
	}
	\label{table_G_result}
	\centering
	\setlength{\tabcolsep}{0.5em}
	\begin{tabular}{|c|m{42em}|c|c|c|}
		
		\hline
		
		\bm{$\epsilon$}	&
		\bfseries Degree Distributions &
		\bm{$R_{\mathsf o }$} &
		\bm{$N_a^*$}  &
		\bm{$R_t^*$} \\
		
		\hline
		\hline
		
		&
		$ \Gamma(x) = 0.2857 x + 0.2857 x^{2} + 0.2286 x^{3} + 0.02857 x^{6} + 0.02857 x^{7} + 0.1429 x^{8} $ &
		&
		&
		\\
		
		0 &
		$\lambda(x) = x $ &
		0.857 &  
		7 &
		0.252 \\
		
		&
		$\rho(x) = x^{13} $ &
		&
		&
		\\

		\hline
		
		&
		$\Gamma(x) = 0.4762 x + 0.4286 x^{4} + 0.09524 x^{10} $ &
		&
		&
		\\
		
		0.1 &
		$\lambda(x) = x $ &
		0.813 &
		7 &
		0.233 \\ 
		
		&
		$\rho(x) =0.0303 x^{2} + 0.09091 x^{7} + 0.06061 x^{8} + 0.09091 x^{9} + 0.1212 x^{10} + 0.303 x^{12} + 0.303 x^{13} $ &
		&
		&
		\\
		
		\hline
		
	\end{tabular}
\end{table*}

\begin{theorem}
	\label{theo_MacLowerBound}
In the absence of the erasure-correcting outer layer,	the fraction of non-recovered bits is lower-bounded by $\Gamma(\epsilon)$.
\end{theorem}

\begin{IEEEproof}
	The collision-resolution layer cannot recover a variable node if all its repetitions on the channel have been erased. All transmissions of a degree $d$ variable node will be erased by the channel with probability $\epsilon^d$. By averaging over $d$, we obtain
	\begin{equation}
		\mathbb{E} \left\lbrace \epsilon^d \right\rbrace 
		= \sum_i{ \Gamma_i \epsilon^i } \\
		= \Gamma(\epsilon).
	\end{equation}
	 Therefore, the probability of erasing all transmissions of a bit is $\Gamma(\epsilon)$. 
\end{IEEEproof}

Also, note the trade-off for the outer code rate: on one hand, a low outer code rate increases the correction capability of the erasure-correcting layer, but on the other hand, it raises the load on the collision-resolution layer, increasing the collisions and the risk of a decoding failure.

\section{Asymptotic Analysis of Decoding Threshold - Short Algebraic Codes} \label{sec_assymptotic_classic}

\subsection{Erasure Transfer Function of MAP Decoder} \label{subsec_erasuretransfer_classic}
In Section \ref{subsec_erasuretransfer_LDPC}, we calculated the equations for erasure transfer functions of time and variable nodes. To incorporate the performance of block MAP decoders in the analysis, we first need to derive their erasure transfer function. More specifically, we are looking for the probability of a decoding failure, given the i.i.d. erasure probability of codeword bits. 

Let $\mathcal{E} \subseteq \{1,2,...,n\}$ be the set of indices of erased elements in a codeword, and let $\bar{\mathcal{E}}$ be its complement. Denote by $\bm{c}_\mathcal{E}$ the sub-vector of erased elements in a codeword $\bm c$, and by $\bm{H}_\mathcal{E}$ the sub-matrix of columns of the parity check matrix $\bm H$ indexed by $\mathcal{E}$. For a linear block code $C(n,k,d\subsc{min})$ with the parity check matrix $\bm{H}$, $\bm{c}$ is a codeword if and only if $\bm{H}\bm{c}^T = \bm 0$, 
\begin{align}
	& \bm H_\mathcal{E}\bm{c}_\mathcal{E}^T   =  \bm H_{ \bar{\mathcal{E}} }\bm c_{ \bar{\mathcal{E}} }^T  \triangleq \bm s^T.  \label{eq_Hepsilon}
\end{align}
If bit erasures happen independently and equiprobably, every sub-codeword $\bm{c}_\mathcal{E}$ satisfying \eqref{eq_Hepsilon} is equally likely. If the system of equations in \eqref{eq_Hepsilon} leads to only one answer $\bm{c}_\mathcal{E}$, then the MAP decoder is successful in retrieving all the erasures.  This is equivalent to $\bm H_\mathcal{E}$ having a full column rank \cite[Ch. 3]{modern_coding_theory}:

\begin{equation}
	\text{rank}\left( \bm H_\mathcal{E} \right) = \left| \mathcal{E} \right|.\label{eq_rank}
\end{equation}
In the sequel we assume that the decoder declares a decoding failure if \eqref{eq_rank} is not satisfied, that is, the decoder does not make any guesses if there is an ambiguity. 
Denote by $P_E(\bm H)$ the ratio of the number of sub-matrices of $\bm H$ with non-complete column rank to the total number of sub-matrices of $\bm H$ with $E$ columns:
\begin{equation}
	P_E(\bm{H}) = \frac{1}{\binom{n}{E}} \times \sum_{ \substack{  \mathcal{E} \subseteq \{1,2,...,n\} \\ |\mathcal{E}|=E }} \mathbbm{1}_{\text{rank}(\bm H_\mathcal{E})<E}, 
	 \label{eq_PEH}
\end{equation}
where $\mathbbm{1}_{\text{test}}$ is an indicator function, which is $1$ when the test is satisfied and zero otherwise.
In fact, $P_E(\bm H)$ is the probability of decoding failure given $E$ erasures. Furthermore, let \PEPin~be the probability of having $E$ erasures in a codeword of length $n$, assuming that the bit erasures are independent and happen with probability $p\subsc{in}$:
\begin{equation}
	\pi_E(p\subsc{in}) = \binom{n}{E} (p\subsc{in})^{E} (1-p\subsc{in})^{n-E}.
\end{equation}
The probability of decoding failure can be written as follows
\begin{align}
	P_e &= \sum_{E = 0}^{n} \pi_E(p\subsc{in}) P_E(\bm H). \label{eq_PE_PH}
\end{align}

\begin{figure}[b!]
	\begin{center}
		\includegraphics[width=\linewidth, height=7cm,keepaspectratio]{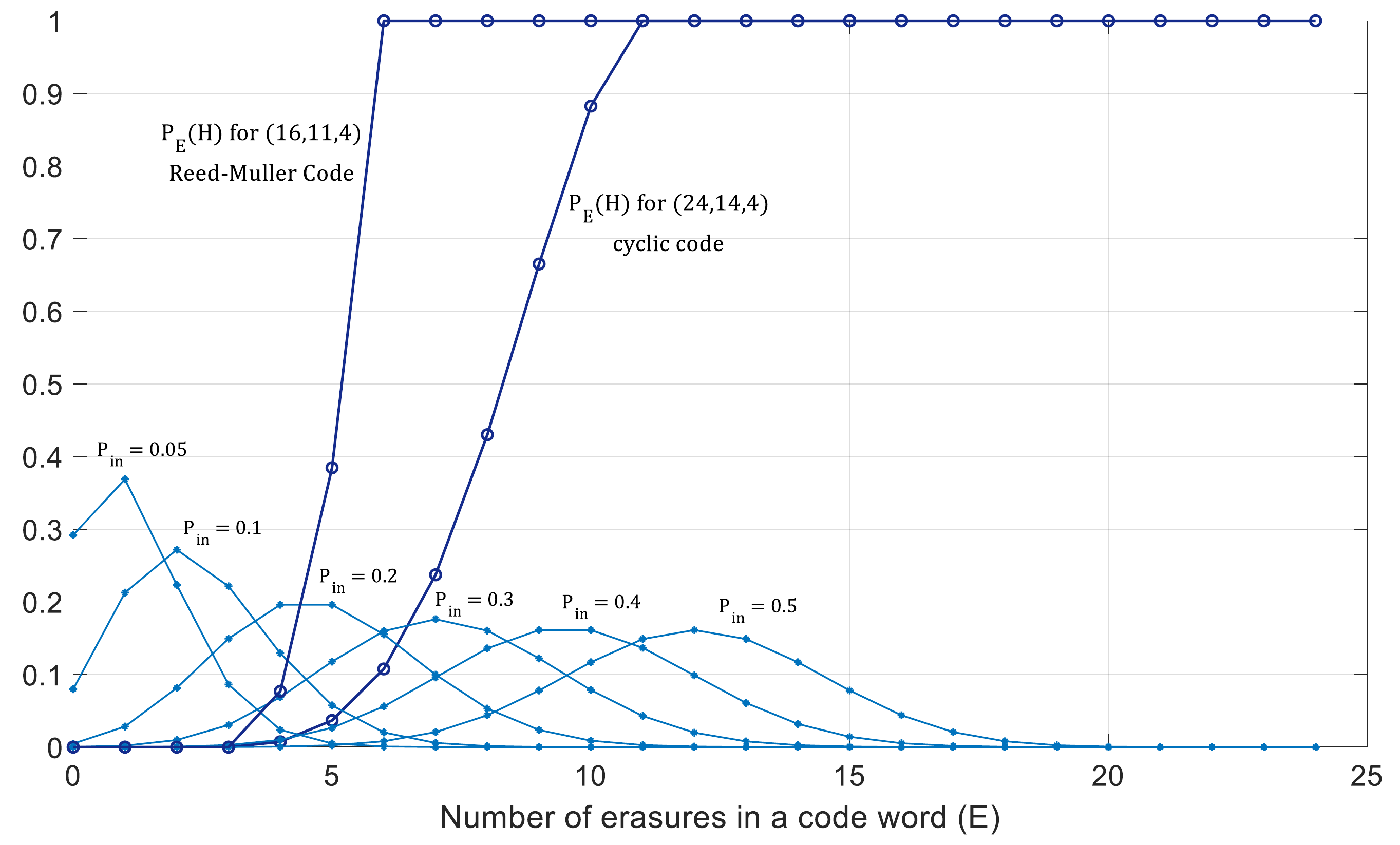}
	\end{center}
	\caption{$P_E(\bm H)$ and \PEPin~for the $(24,14,4)$ cyclic code with the generator polynomial $1+x^4+x^6+x^{10}$ and $(16,11,4)$ Reed-Muller code.}
	\label{fig_PEH_PEPin}
\end{figure}

Having $P_E(\bm H)$ values of a linear block code for $0 \leq E \leq n$, one can determine the exact performance of the code under i.i.d. erasures using \eqref{eq_PE_PH}. In fact, $P_E(\bm H)$  characterizes the erasure correction capability of the code, and \PEPin~represents the erasure imposed by the channel. Fig. \ref{fig_PEH_PEPin} shows $P_E(\bm H)$ coefficients for the $(24,14,4)$ cyclic code with the generator polynomial $1+x^4+x^6+x^{10}$ and the $(16,11,4)$ Reed-Muller (RM) code, as well as \PEPin~for different values of $p\subsc{in}$. It is evident from \eqref{eq_PE_PH} that when the overlap of non-zero values of $P_E(\bm H)$ and \PEPin~(as functions of $E$) is smaller, the erasure correction performance of the code is stronger.

\begin{figure}[b!]
	\begin{center}	
		\includegraphics[width=\linewidth, height=8cm,keepaspectratio]{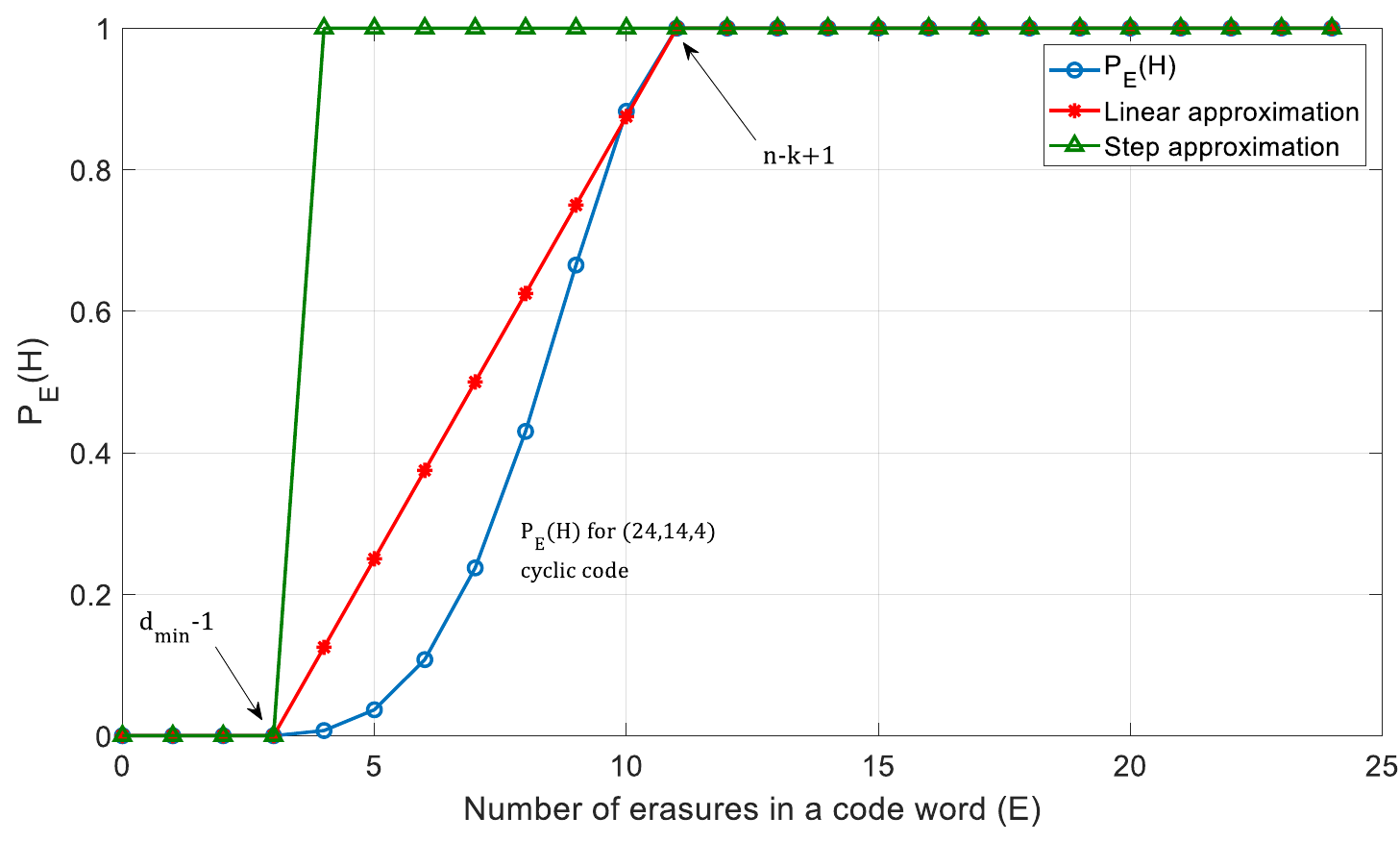}
	\end{center}
	\caption{Exact $P_E(\bm H)$ and its step and linear approximations.}
	\label{fig_approx}
\end{figure}

When Gaussian elimination is used to exactly compute $P_E(\bm H)$, that computation is only feasible for short codes as it demands $O(2^n n^3)$ operations, with $O(n^3)$ operations to find the rank of $2^n$ matrices corresponding to each erasure pattern. However, we can consider a few simplifying facts. First, every linear block code is capable of correcting all erasure patterns of weight $d\subsc{min} - 1$ or less. In addition, the rank of a parity check matrix (and all its sub-matrices) is at most $n-k$. Therefore,

\begin{align}
	P_E(\bm H) &= 0, \hspace{2em}  E < d\subsc{min} \label{eq_PEH_low} \\
	P_E(\bm H) &= 1, \hspace{2em}  E > n - k  \label{eq_PEH_high}.
\end{align}
As a result, we only need to compute $P_E(\bm H)$ for $d\subsc{min} \leq E \leq n-k$, which is still computation-intensive for long or low-rate codes. An alternative approach is to approximate $P_E(\bm H)$. Equation \eqref{eq_step} below demonstrates a step approximation which assumes the code can only correct erasure patterns of weight less than \dmin. In fact, this approximation leads to the exact failure probability of a bounded distance decoder, and to an upper bound for a MAP decoder.
\begin{equation} 
	\label{eq_step}
	P_E(\bm H) = 
	\begin{cases}
		0, \hspace{2em} 	& E 	< d\subsc{min} \\
		1, \hspace{2em} 	& E \geq d\subsc{min}
	\end{cases}
\end{equation}
In the same direction, we can consider a linear approximation of $P_E(\bm H)$ for  $d\subsc{min} \leq E \leq n-k$: 
\begin{equation} 
	\label{eq_linear}
	P_E(\bm H) = 
	\begin{cases}
		0, & E < d\subsc{min} \\
		\frac{E-(d\subsc{min}-1)}{(n-k+1)-(d\subsc{min}-1)}, \hspace{1em} d\subsc{min} \leq & E \leq n-k \\
		1, & E > n-k.
	\end{cases}
\end{equation}
The approximations are more accurate for codes whose minimum distance approaches $n-k$. In fact, the approximation is exact when the code is MDS.
Fig. \ref{fig_approx} compares the exact values of $P_E(\bm H)$ for a (24,14,4) cyclic code with step and linear approximations in \eqref{eq_step} and \eqref{eq_linear}, respectively.

\subsection{Tracking Decoder Erasure Probability} \label{subsec_trackdecoder_classic}
In the previous section, we characterized the probability of block MAP decoding failure. We can now track the evolution of erasure probability of messages passed along the edges of the Tanner graph in the proposed two-layer decoding architecture of Fig. \ref{fig_decoder_b}, for a fixed $N_a$ and $\epsilon$. Fig. \ref{fig_de_b} depicts the messages passed in one iteration. By expressing the erasure probability of each message in terms of the erasure probability of its generating messages, we can reach a set of recursive equations for $P_{ct}^{(\ell)}$. Using variable and time node erasure transfer functions in \eqref{eq_vnode_tf} and \eqref{eq_tnode_tf}, we have 

\begin{align}
	P_{tv}^{(\ell)} &= 1- (1-\epsilon) \psi \left( 1- \frac{P_{ct}^{(\ell)}}{2} \right)  \label{eq_rec_classic_1} \\
	P_{vc}^{(\ell)} &= \gamma \left( P_{tv}^{(\ell)} \right) \label{eq_rec_classic_2}
\end{align}

From \eqref{eq_PE_PH}, we have

\begin{align*}
	P_{ct}^{(\ell)} &= \Pr \left( m_{ct}^{(\ell)} = e \right) \\
					&= \Pr \left( m_{vc}^{(\ell-1)} = e , \text{decoding failure} \right) \\
					&= P_{vc}^{(\ell-1)}
					\sum_{E=1}^{n} \binom{n-1}{E-1} {P_{vc}^{(\ell-1)}}^{E-1} \left( 1-P_{vc}^{(\ell-1)} \right) ^{n-E} P_E(\bm H) \numberthis \label{eq_rec_classic_3}
\end{align*}
Since at the beginning of the decoding, every variable node is unknown and all edges carry erasure messages, the initial value for the recursion is

\begin{equation}
	P_{ct}^{(0)} = 1 \label{eq_rec_classic_4}.
\end{equation}
Unlike Section \ref{sec_assymptotic_LDPC} where large code lengths allow for asymptotic analysis, in case of short codes for density evolution equations \eqref{eq_rec_classic_1} to \eqref{eq_rec_classic_3} to be valid in each frame, a large number of active users $N_a$ is assumed. In this case, as seen in the decoder graph of Fig.~\ref{fig_decoder_b}, the overall size of the decoding block ($N_a n$ and $T$) remains large. 

\subsection{Optimizing Code and Repetition Rate Distribution} \label{subsec_optimization_classic}

Using the recursive set of equations in \eqref{eq_rec_classic_1}-\eqref{eq_rec_classic_4}, now we can compute the user rate threshold $R_t^*$ corresponding to a repetition rate distribution $\{\Gamma\}$, parity check matrix $\bm H$, the maximum number of active transmitters $N_a^*$, and channel erasure probability $\epsilon$. Considering the thresholding behavior of the decoder, if for the user rate $R_t$ the limit $P_{ct}^{(\infty)} = \lim_{\ell \to \infty} P_{ct}^{(\ell)} > 0$, where $P_{ct}^{(\ell)}$ is defined in \eqref{eq_rec_classic_4}, then $R_t^*<R_t$. On the other hand, a zero value of $P_{ct}^{(\infty)}$ implies that $R_t^* \geq R_t$. Therefore, we can perform a binary search on $R_t^*$ to calculate the user rate threshold for a fixed $N_a^*$ and $\epsilon$.

Now we can search for the pair of optimized code and repetition rate distributions; i.e., the optimized code is selected from the ensemble of $\mathscr{C} = \left\lbrace \mathcal{C}_1(n_1,k_1,d_1),...,\mathcal{C}_{|\mathscr{C}|}(n_{|\mathscr{C}|},k_{|\mathscr{C}|},d_{|\mathscr{C}|}) \right\rbrace $ along with an optimized degree distribution $\{\Gamma\}$. Furthermore, $\mathscr{C}$ could be part of a code family, e.g., Reed-Muller codes, Hamming codes, or Expander codes, or it could be a set of arbitrary binary linear block codes. We seek to solve the following optimization. 

\begin{alignat*}{4}
& \operatorname*{maximize}_{\{ \Gamma\}, \hspace{1pt } \mathcal{C}(n,k,d)} \quad && R_t^*(N_a^*,\epsilon) \numberthis \label{eq_opt_classic} \\
& \text{subject to:} 
&& \mathcal{C} \in \mathscr{C} \\
&	&& R_t^* \leq R_\mathcal{C} / \Gamma'(1) \\
&	&& \Gamma_i = 0, &&& i > d\subsc{max} \\
&	&& \Gamma_i \in [ 0,1 ], &&& i \leq d\subsc{max} \\
&	&& \textstyle \sum_i \Gamma_i = 1
\end{alignat*}
Similar to Section \ref{subsec_optimization}, to this end we use differential evolution \cite{differential_evolution}.
Table \ref{table_opt_classic} shows the optimized codes and degree distributions using the linear approximation \eqref{eq_linear} and Reed-Muller codes up to length 128 as the ensemble $\mathscr{C}$.
\begin{table}[htb!]
    \centering
	\caption{Optimized codes and degree distributions for Reed-Muller codes up to length 128 and $\epsilon = 0.1$, $d\subsc{max} = 13$.}
	\label{table_opt_classic}
	\begin{tabular}{|c|l|c|}	
		\hline
		 $\bm{N_a^*}$ & \bfseries Optimized Degree Distribution & \bfseries Selected Code \\
		\hline
		\hline
		6  & $\Gamma(x) = 0.4063 x^{3} + 0.5625 x^{4} + 0.03125 x^{13}$ & RM (7,5)\\
		\hline
		31 & $\Gamma(x) = 0.9091 x^{4} + 0.09091 x^{13}$ &  RM (6,5)\\
		\hline
	\end{tabular}
\end{table}

\section{Performance Evaluation} \label{sec_results}
In this section, we present numerical and simulation results to analyze the performance of our  two-layer coded channel access scheme over a binary adder channel with erasures. We first consider the multiple access scenario (TCMA), where the number of active users $N_a$ is known both at the transmitters and at the receiver. Next, we consider the random access scenario, where the number of active users is random in each frame and only known at the receiver (TCRA). For comparison, we consider several performance limits and a benchmark scheme that are described in the sequel.

In simulating the outer layer in the proposed scheme with LDPCs, the LDPC codes are constructed randomly from optimized degree distributions by avoiding cycles of size 4. More advanced code construction methods can be considered for practical implementation. For the short algebraic block codes, we use the family of Reed-Muller codes as the search space $\mathscr{C}$ in the optimization problem \eqref{eq_opt_classic}. Reed-Muller family \cite{muller, reed} offers a class of highly structured linear block codes with predefined minimum distance. This is important in the current setting, since it allows us to characterize the performance of the code family using the approximations \eqref{eq_step} and \eqref{eq_linear} simply by the code parameters $m^{(RM)}$ and $r^{(RM)}$.

\subsection{Multiple Access Channel}
\begin{figure}[b!]
	\begin{center}
		\includegraphics[width=\linewidth, height=9cm,keepaspectratio]{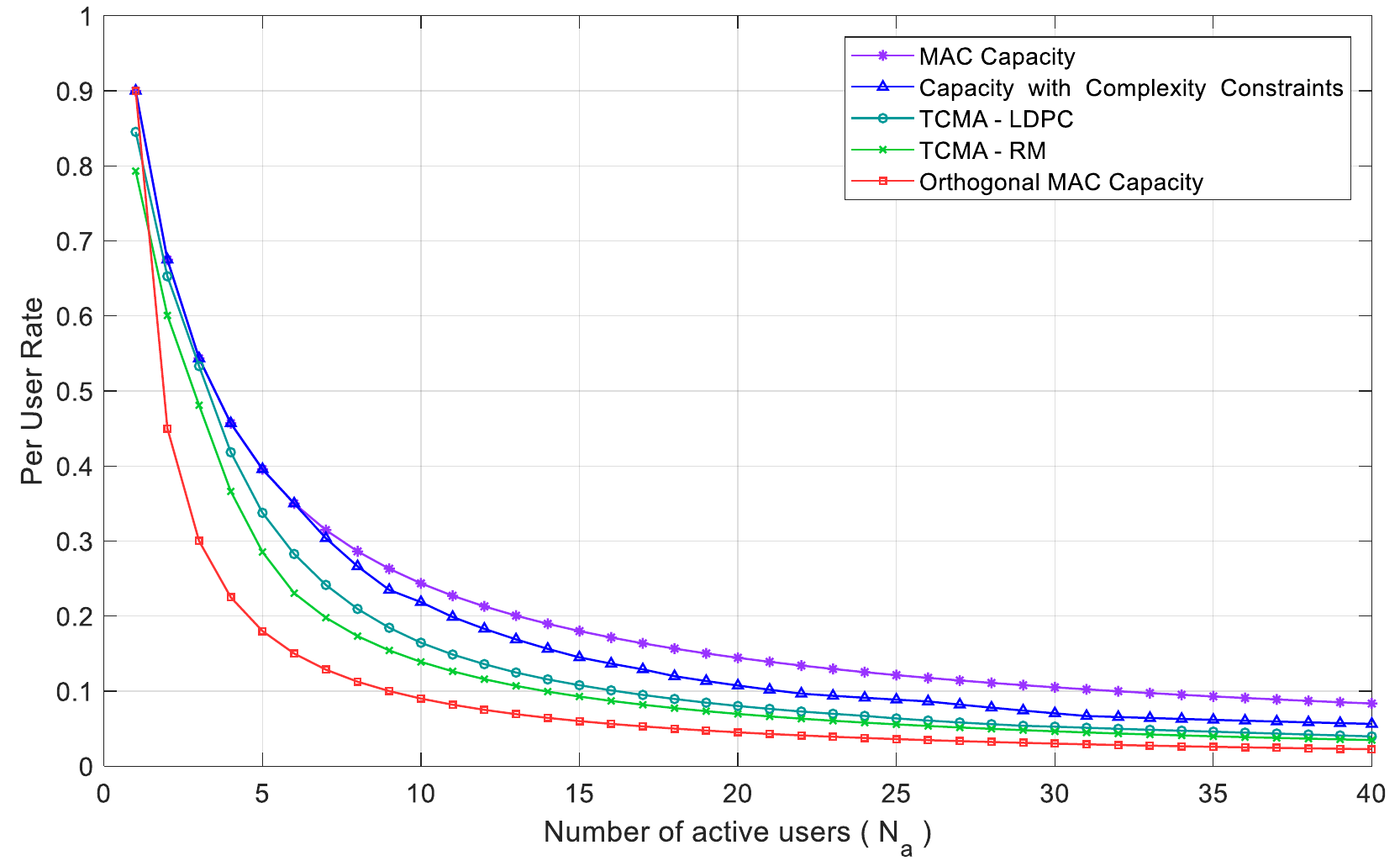}
	\end{center}
	\caption{Transmission rate per user (threshold from analysis) for the proposed Two-layer Coded Multiple Access scheme using LDPC codes and RM codes for $\epsilon = 0.1$ and $d\subsc{max}=13$.
	}
\label{fig_Rt_Na}
\end{figure}

Fig. \ref{fig_Rt_Na} depicts the rate per user (b/s/Hz) of the proposed TCMA with LDPC codes as a function of the number of active users over the binary adder channel with erasure probability $\epsilon=0.1$. The curves are obtained by solving the design optimization problem \eqref{eq_opt} for each $N_a^*$ and $d\subsc{max}=13$. 
For comparison, we also present the per-user capacity of the multiple access channel \eqref{EqCNa}, and the per-user capacity with orthogonal transmission, given by 
\begin{align}
\frac{C_1}{N_a} = \frac {1 - \epsilon}{N_a}. \label{eq_ort}
\end{align}
As evident in this figure, the proposed scheme noticeably outperforms an ideally coded orthogonal transmission scheme. Specifically with five users, compared to such an ideal scheme, the proposed TCMA-LDPC reduces the gap to MAC capacity by approximately two thirds.

Of interest is also a comparison with a complexity constrained capacity of binary adder MAC with erasures in which user transmission patterns are constrained to satisfy the same time-node degree distribution $\psi(x) = \sum_{k = 1}^{K} \psi_k x^{k}$ as the one that optimizes \eqref{eq_opt}. The complexity-constrained capacity is computed as 
\begin{equation}
 \frac 1 {N_a} \sum_{k = 1}^{K} \psi_k \, C_{k}, \label{eq_comp}
\end{equation}
where $C_{k}$ is given in \eqref{EqCNa}.
Of course, the orthogonal per-user MAC capacity \eqref{eq_ort} is a special case of \eqref{eq_comp} with $\psi(x) = x$. Another important special case is the capacity of the MAC channel with $\leq K$-fold collisions allowed in each time-slot. In that case, the per-user capacity is given by $\frac{C_{K}}{N_a}$, which corresponds to $\psi(x) = x^K$. We see from Fig. \ref{fig_Rt_Na} that taking decoding complexity considerations into account bridges part of the gap to the MAC capacity. We stress that the performance of the proposed TCMA-LDPC is obtained with a suboptimal decoder whose complexity grows only linearly with the number of users, code length and node degrees.  

Also in this figure, the rate per user performance of the proposed TCMA with Reed-Muller codes is depicted. The performance is obtained by solving the design optimization problem \eqref{eq_opt_classic} for each $N_a^*$ and $\epsilon=0.1$ and $d\subsc{max}=13$. The TCMA-RM provides a competitive rate performance in comparison, while maintaining a low application level delay and data rate (payload size) requirement due to its short code block length. These characteristics are of substantial interest in next generation internet of things and machine to machine applications.

\subsection{Random Access Channel}

For different maximum number of active transmitters $N_a^*$, we designed optimized TCRA codes with LDPC codes (Section \ref{sec_assymptotic_LDPC}) and the family of Reed-Muller codes (Section \ref{sec_assymptotic_classic}). Translating the maximum number of active transmitters into target outage probabilities for a Poisson distribution of number of active users, we can evaluate the performance of the system for different average number of active users, $Np_a$ (for Poisson distribution, $P\subsc{outage}$ is a function of the average number of active users $Np_a$ only).

\begin{figure}[tb!]
	\begin{center}
		\includegraphics[width=\textwidth, height=10cm,keepaspectratio]{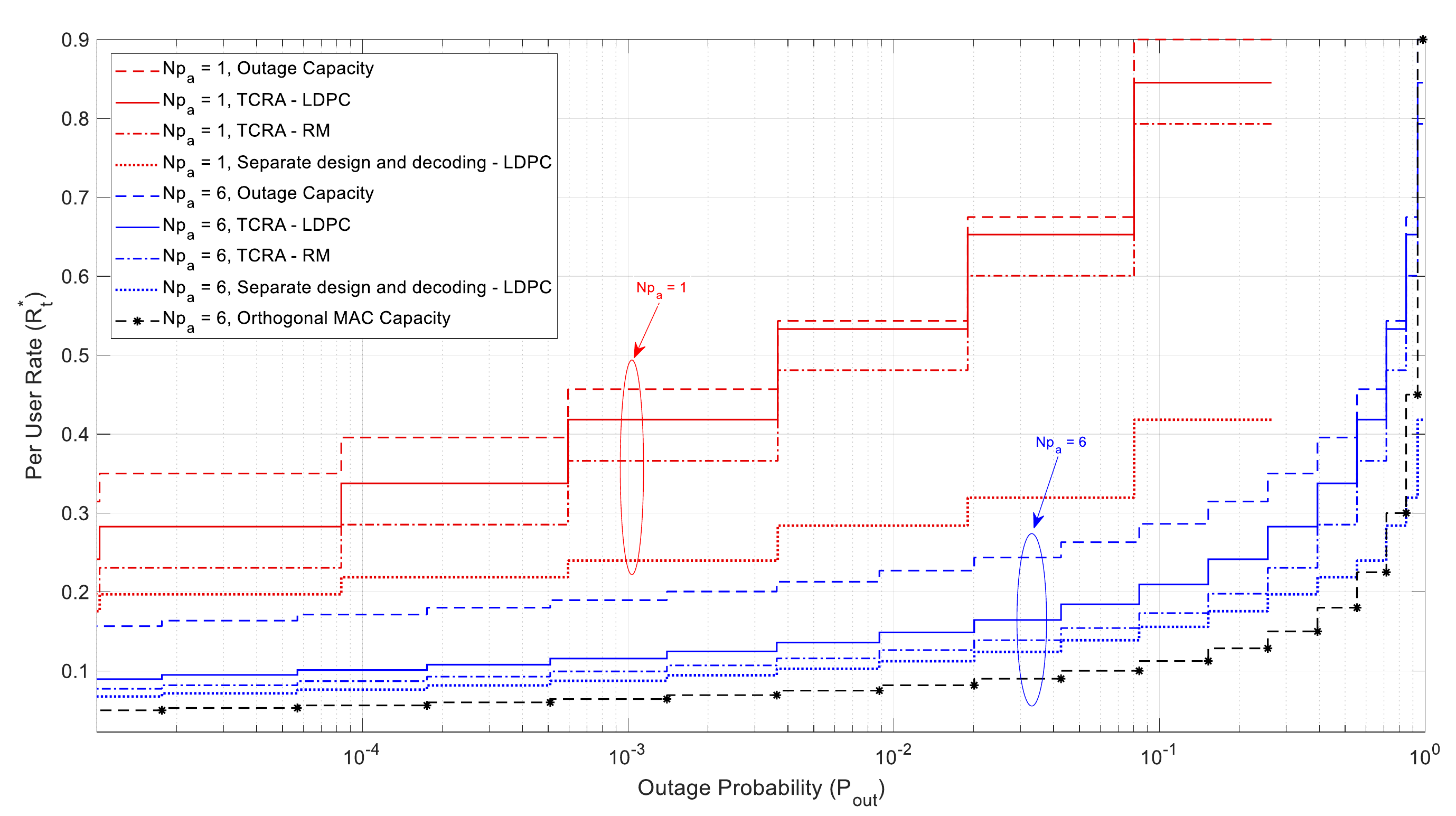}
	\end{center}
	\caption{Transmission rate per user (threshold from analysis) for Two-layer Coded Random Access scheme using LDPC and Reed-Muller codes, separately designed/decoded system, and outage capacity for $\epsilon=0.1$ and $d\subsc{max} = 13$.  The black starred curve shows the orthogonal signaling capacity for $Np_a = 6$.}
	\label{fig_cap}
\end{figure}

Fig. \ref{fig_cap} demonstrates the optimized user rate threshold $R_t^*$ for TCRA as a function of the outage probability using the LDPC codes and the family of Reed-Muller codes with $m^{(RM)} < 5$, for $\epsilon=0.1$ and $d\subsc{max} = 13$, along with the corresponding outage capacity curves (without complexity constraints). The exact performance of the RM codes ($P_E(\bm H)$) has been used to solve \eqref{eq_opt_classic} and design the system.

The proposed design noticeably outperforms the outage capacity of orthogonal signaling (computed by letting $C_{Na} = C_1 = 1 - \epsilon$ in \eqref{eq_Cout}), confirming that indeed allowing collisions in this framework is beneficial. Furthermore, we compare the performance of our design with a benchmark system in the spirit of \cite{CRDSA}-\cite{rateless}, in which the inner layer is optimized for a non-erasure channel and the outer layer is an LDPC code designed for the BEC($\epsilon$). 
Unlike the proposed TCRA system in which the inner layer and the outer layer cooperate by iteratively passing messages back and forth to achieve a joint resolution of collisions and erasures, the two layers in the benchmark scheme operate in tandem.
The proposed scheme noticeably outperforms the benchmark. 
Also evident is a rather graceful trade-off of quality of service (quantified by probability of outage) and the throughput performance (achievable rate). A more stringent QoS constraint imposes a greater toll on throughput. 

\begin{figure}[tb!]
	\begin{center}
		\includegraphics[height=10cm,keepaspectratio]{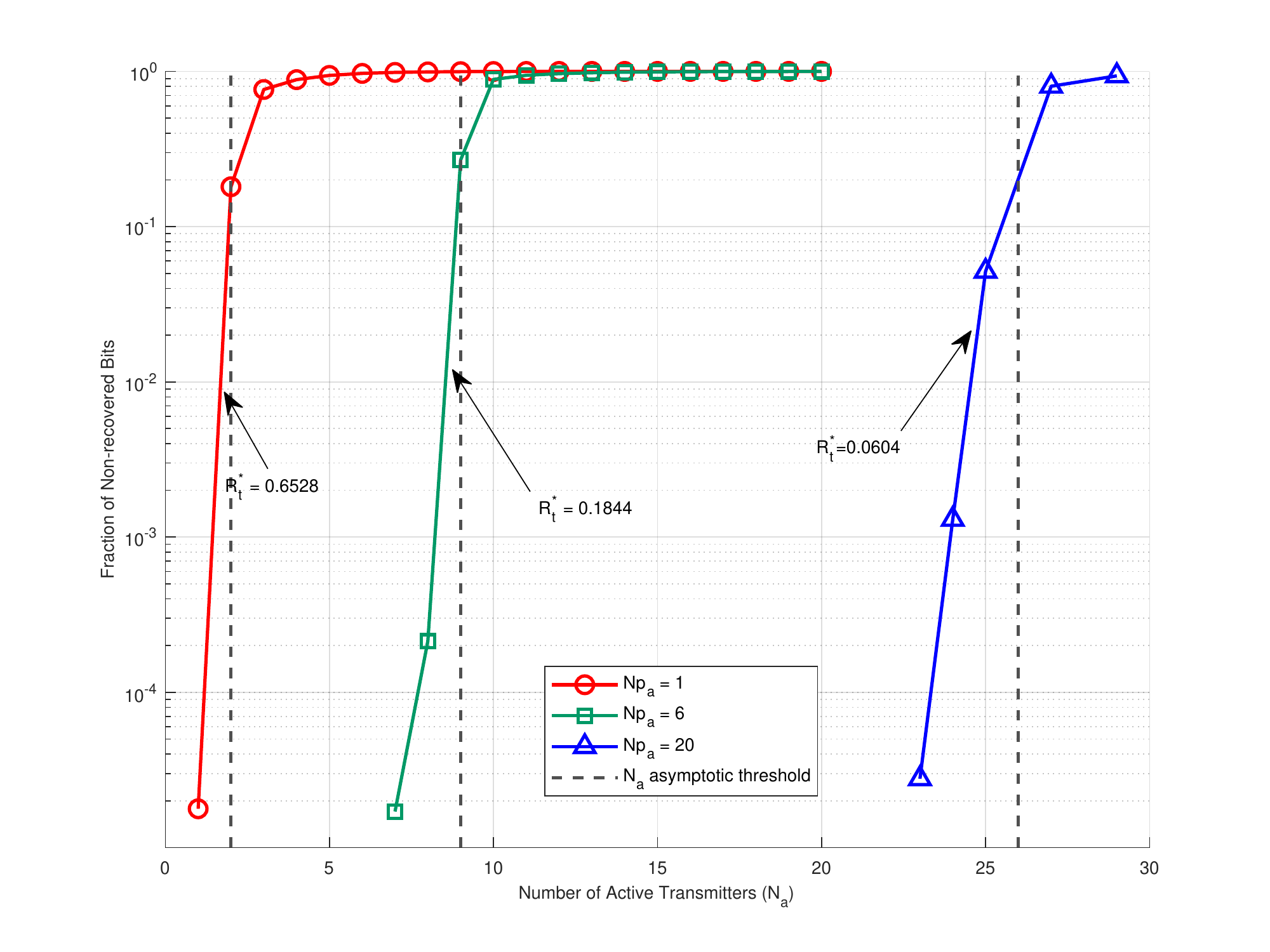}
	\end{center}
	\caption{Monte Carlo simulation showing the fraction of non-recovered bits as a function of the number of active transmitters for the Two-Layer Coded Random Access system with $\epsilon=0.1$, $d\subsc{max} = 13$, $P\subsc{outage} = 0.1$, at user block length of 1000 bits.
	}
	\label{fig_performance}
\end{figure}

Fig. \ref{fig_performance} shows the simulation results for the performance of the proposed TCRA-LDPC scheme with the efficient decoder presented in Section \ref{subsec_decrx} as a function of the actual number of active transmitters. Although the curves in Fig. \ref{fig_performance} are evaluated for limited code and frame lengths, in line with our asymptotic analysis, the system exhibits a threshold effect at the designed maximum number of active senders, $N_a^*$.

\begin{figure}[tb!] 
	\begin{center}
		\includegraphics[height=12cm,keepaspectratio]{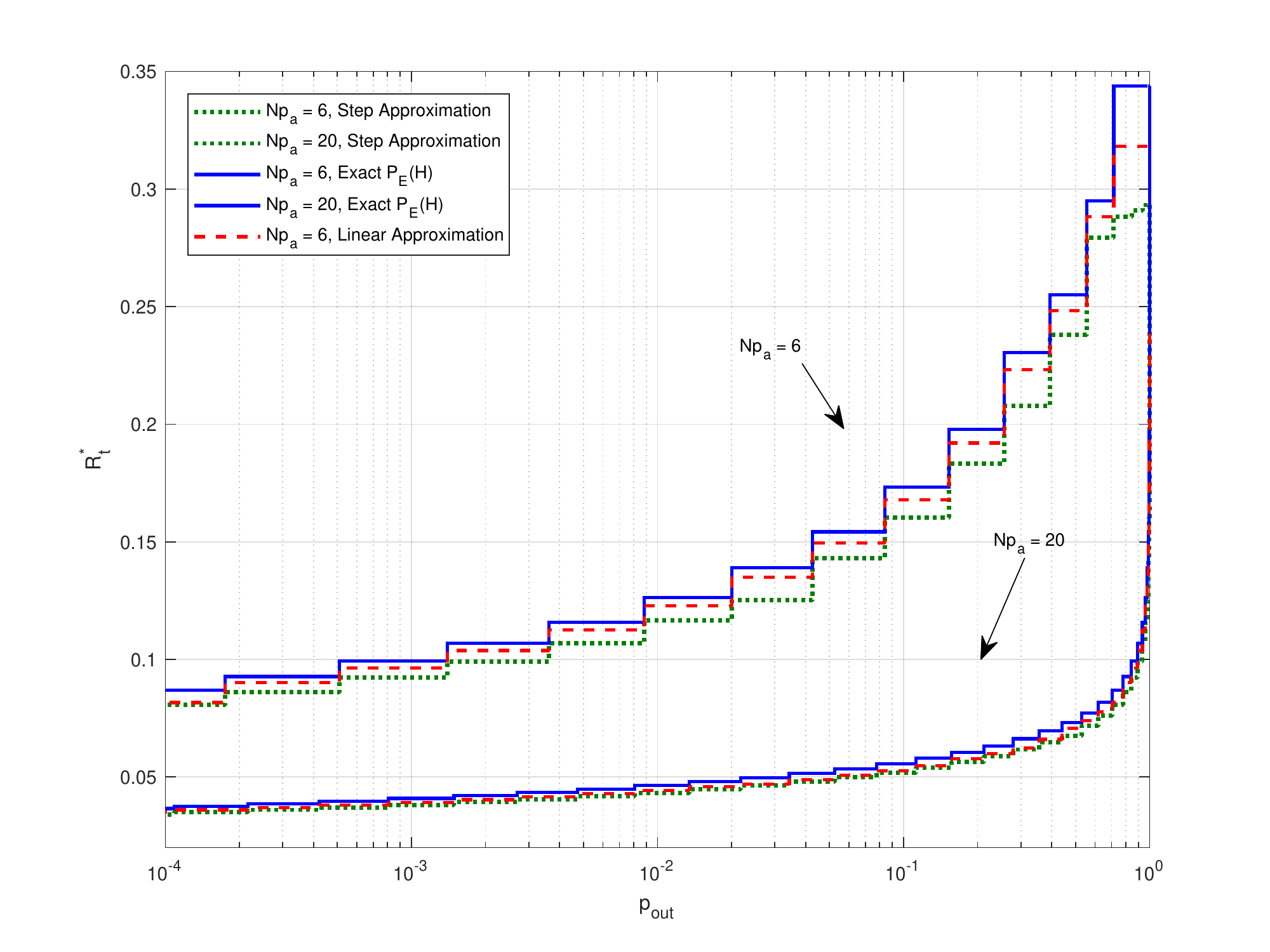}
	\end{center}
	\caption{Optimized user rate thresholds for TCRA-RM obtained by the exact and approximated values of $P_E(\bm H)$ as a function of probability of outage with $\epsilon=0.1$ and $d\subsc{max} = 13$.}
\label{fig_approximations}
\end{figure}

Fig. \ref{fig_approximations} compares the optimized user rate threshold of TCRA-RM with $m^{(RM)}<5$ designed with the exact and approximated values of $P_E(\bm H)$ for $\epsilon=0.1$ and $d\subsc{max} = 13$. As expected, compared to the step approximation, the optimized rates obtained based on linear approximation are closer to those based on exact values of $P_E(\bm H)$. In fact, the codes selected in the optimization process for linear approximation and exact $P_E(\bm H)$ are the same. Also, the optimized degree distributions resulted from both approaches are very similar. Nonetheless, as discussed in Section \ref{sec_assymptotic_classic}, the design of TCRA with algebraic short codes are computationally much more efficient when the code performance is approximated as prescribed. 

\section{Practical Remarks}

In this section, we briefly comment on some practical issues on the operation of TCRA. Briefly, here is how the system operates: The receiver announces the start of a frame via a beacon signal in which it also informs the transmitters of the current $N_a^*$. This is estimated based on a specific target outage probability and some model of user activity as discussed in the beginning of this section. The transmitters based on the code choices and designs agreed upon during the setup phase, would then know the code in the outer layer (fixed for all users in the current setup), and the degree distribution in the inner layer and the frame length. Each transmitter will then use this information to randomly place the coded bits over the $T$ time slots. This may be accomplished using a properly seeded random generator. For the receiver to decode the messages, it needs to know the inner layer codes (active users/seeds) to obtain the full decoding graph. In principle, there are several ways to this end: (1) The receiver may learn this directly from the received sequence via a joint user (seed) identification and decoding approach. This is in the same spirit as the rich literature on blind joint CDMA code detection and decoding, e.g., {\cite{burel2001detection}}, {\cite{nzeza2006parallel}}. (2) Each user may embed a user ID (with one to one correspondence with the seed) in each time slot transmission. This is the approach suggested in \cite{IRDSA} for the packetized contention resolution random access. (3) Alternatively, we may consider a one-way \emph{check-in procedure} at the beginning of the transmission frames, wherein the users announce their \emph{intention-to-transmit} within the upcoming frame. This can be carried out using contention-based \cite{AFC} or non-contention-based RA schemes. One may argue that such a check-in procedure can be exploited by the receiver to coordinate the users. In this scenario, the system would operate in the TCMA mode as opposed to the TCRA mode. However, we emphasize that there is no advantage in attempting to schedule the users since the proposed schemes favourably exploit the collisions for improved performance as presented in Figs. {\ref{fig_Rt_Na}} and {\ref{fig_cap}}. While, in the above performance evaluation, we assumed perfect knowledge of the set of active users in each frame at the receiver, further investigation of this issue remains for future research.

\section{Conclusions} \label{sec_conclusion}

We introduced and analyzed a two-layer coded channel access framework for reliable communications over erasure adder multiple access channels, which enables joint resolution of user collisions and erasure correction. The  outer  layer deals with erasures, while the inner layer deals with collision resolution. Both layers are optimized jointly. The decoding  process  proceeds  iteratively,  with  both  layers  exchanging  messages  back  and  forth.

Both a multiple access scenario where the number of users are known at the transmitters and the receiver, and a random access scenario, where this information is only available at the receiver, were considered. In the proposed two-layer coded channel access framework, we presented density evolution analysis for code optimization in cases where the outer layer is a long LDPC code or a short (algebraic) code.  

The results with long LDPC codes demonstrated a superior performance for the proposed scheme in comparison to the capacity of an equivalent orthogonal MAC, and the related prior art. The proposed message passing decoder offers a linear complexity as a function of number of users and node degrees and approaches the capacity of the corresponding channels. With short codes from the Reed Muller family, the proposed scheme provides a competitive performance, while maintaining a low application level delay and data rate (payload size) requirement due to its short code block length. These characteristics are of substantial interest in next generation internet of things and machine to machine applications.

In this work, we considered binary erasure adder access channel. This channel model also serves as a simpliﬁed model for binary signaling of multiple users over a noisy channel, where a received signal with low reliability may be interpreted as erasure. The simplicity of the model allowed us to develop, investigate and gain insight into the elements of a high performance and scalable coded access scheme with collision resolution. Future research could examine code design with other iterative decoding schemes, and possibly over other types of multiple access communications channels. Specifically, of prime research interest are practical issues such as  synchronization in presence of symbol collisions \cite[Appendix B]{IRDSA} and design and decoding of TCRA over Gaussian channel and over fading channels. Another line of research is to aim for a more accurate analysis of the proposed coded access schemes with short algebraic codes. For example, if the weight enumerator of the linear block code is known as in the case of some algebraic codes {\cite{liva2013bounds}},  one may approximate the outer layer block error probability using a union bound.

\bibliographystyle{IEEEtran}
{\footnotesize
\bibliography{references}}

\vspace{10pt}

\end{document}